\documentclass[11pt,twoside]{article}
\usepackage{macro-fsut-eng}
\usepackage{graphicx}

\usepackage[T1]{fontenc} 

\usepackage{latexsym}
\usepackage{verbatim}
\usepackage{amssymb}

\begin{document}

\vskip 1.0cm
\markboth{E. F. Eiroa and C. M. Sendra}{Regular phantom black holes as gravitational lenses}
\pagestyle{myheadings}

\vspace*{0.5cm}
\title{Regular phantom black holes as gravitational lenses}

\author{Ernesto F. Eiroa$^{1,2}$ and Carlos M. Sendra $^{1,2}$}
\affil{$^1$Instituto de Astronom\'{\i}a y F\'{\i}sica del Espacio, Buenos Aires, Argentina\\
$^2$Departamento de F\'{\i}sica, FCEN-UBA Buenos Aires, Argentina}

\begin{abstract}
The distortion of the spacetime structure in the surroundings of black holes affects the trajectories of light rays. As a consequence, black holes can act as gravitational lenses. Observations of type Ia supernovas, show that our Universe is in accelerated expansion. The usual explanation is that the Universe is filled with a negative pressure fluid called dark energy, which accounts for 70\% of its total density, which can be modeled by a self-interacting scalar field with a potential. We consider a class of spherically symmetric regular phantom black holes as gravitational lenses. We study large deflection angles, using the strong deflection limit, corresponding to an asymptotic logarithmic approximation. In this case, photons passing close to the photon sphere of the black hole experiment several loops around it before they emerge towards the observer, giving place to two infinite sets of relativistic images. Within this limit, we find analytical expressions for the positions and the magnifications of these images. We discuss the results obtained and make a comparison with the Schwarzschild and Brans-Dicke solutions for the case of the galactic center supermassive black hole.
\end{abstract}

\section{Introduction}
\label{sec:introduction}
The study of gravitational lensing by black holes has received a boost (Virbhadra et al. 2000) due to the evidence of the presence of supermassive black holes at the center of galaxies, including ours. For these objects, two sets of relativistic images are formed when the light rays pass close to the photon sphere, for which the strong deflection limit is adopted. This approximation was found for any spherically symmetric object with a photon sphere (Bozza 2002). Many works considering strong deflection lenses are found in the literature for different types of black holes. It is thought that observations of the optical effects associated with these objects will be possible in the near future (Eiroa 2012).

The accelerated expansion of the Universe can be explained by the existence of dark energy as the prevailing component (see, for example, Bamba et al. 2012). This can be in the form of phantom energy if $\omega<-1$ in the equation of state $p=\omega\rho$, and can be modeled by a self-interacting scalar field with a potential. Withing this context, regular black hole and wormhole phantom solutions were found (Bronnikov et al. 2006); also phantom dilaton black holes were recently studied as gravitational lenses (Gyulchev et al. 2013; Eiroa et al. 2013).

\section{Regular phantom black hole}
\label{sec:pbh}
We consider the following spherically symmetric geometry, which is a solution of the Einstein equations with a scalar field possessing a negative kinetic term and a potential (Bronnikov et al. 2006):
\begin{equation}
ds^2=-A(r)dt^2+B(r)dr^2+C(r)d\Omega^2,
\end{equation}
where
\begin{equation}
A(r)=B(r)^{-1}=1+\frac{3m r}{b^2}+(r^2+b^2)\left[\frac{c}{b^2}+\frac{3m}{b^3}\tan^{-1}\left(\frac{r}{b}\right)\right], \nonumber
\end{equation}
\begin{equation}
C(r)=r^2+b^2,
\label{m2}
\end{equation}
with $c$, $m$, and $b>0$ constants. The solution is regular everywhere, i.e. free from curvature singularities; $b$ is the scale of the scalar field, and $m$ can be interpreted as the mass. In the particular case that $c=-3m\pi/2b$, the metric becomes asymptotically flat. We have a black hole solution for $m>0$, with a Killing horizon $r_h$. In this case, the region corresponding $r>r_h$ is asymptotically flat, and the one with $r<r_h$ is asymptotically de Sitter. This solution is stable for $b=3m\pi/2$, for which $r_h=0$ (Bronnikov et al. 2012). It is convenient to adimensionalyze the metric with the mass, by defining
\begin{figure}
\begin{center}
\hspace{0.25cm}
   \includegraphics[width=7.5cm]{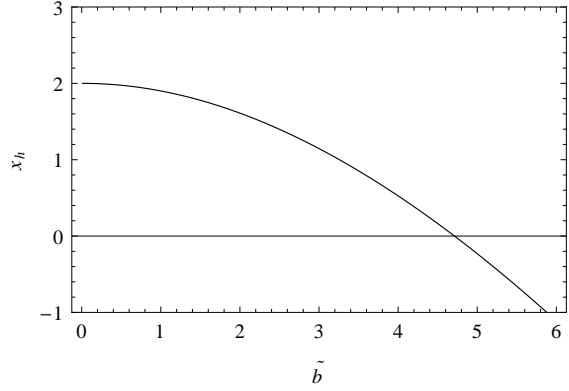}
\caption{Horizon radius $x_h=r_h/m$ as a function of $\tilde{b}=b/m$.}
\label{xh}
\end{center}
\end{figure}
\begin{equation}
x=r/m, \hspace{0.4cm} T=t/m, \hspace{0.4cm} \tilde{b}=b/m,
\end{equation}
and adopting the flatness condition, so the black hole solution has the form
\begin{equation}
ds^{2}=-A(x)dT^{2}+B(x)dx^{2}+C(x)(d\theta^{2}+\sin^{2}\theta d\phi^{2}),
\end{equation}
\begin{equation}
A(x)=B(x)^{-1}=1+\frac{3x}{\tilde{b}^2}+\frac{3}{\tilde{b}}\left(1+\frac{x^2}{\tilde{b}^2}\right)\left[-\frac{\pi}{2}+\tan^{-1}\left( \frac{x}{\tilde{b}}\right)\right],
\end{equation}
\begin{equation}
C(x)=x^2+\tilde{b}^2.
\end{equation}
The adimensionalyzed radius of the horizon $x_h$, corresponding to the root of $A(x)$, is a decreasing function of $\tilde{b}$, shown in Fig. \ref{xh}. The radius of the photon sphere $x_{ps}$ is given by the largest positive solution of the equation
\begin{equation}
A'(x)C(x)=C'(x)A(x),
\end{equation}
where the prime represents the derivative with respect to $x$. For the phantom black hole, corresponds to the constant value $x_{ps}=3$.

\section{Strong deflection limit}
The deflection angle for a photon coming from infinity, as a function of the closest approach distance $x_0$, is given by (Virbhadra et al. 1998)
\begin{equation}
\alpha(x_0)=I(x_0)-\pi,
\end{equation}
where
\begin{equation}
I(x_0)=\int^{\infty}_{x_0}\frac{2\sqrt{B(x)}dx}{\sqrt{C(x)}\sqrt{\frac{C(x)}{C(x_0)}\frac{A(x_0)}{A(x)}-1}}.
\label{I0}
\end{equation}

\begin{figure}
\begin{center}
 \hspace{0.25cm}
   \includegraphics[width=7.5cm]{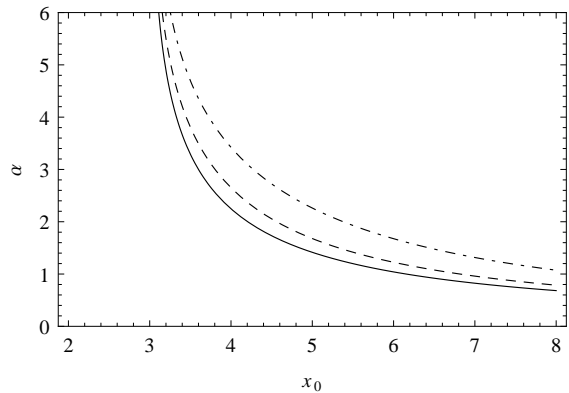}
\caption{Deflection angle $\alpha$ as a function of the closest approach distance $x_0$ for $\tilde{b}=1$ (full line), $\tilde{b}=3$ (dashed line), and $\tilde{b}=6$ (dashed-dotted line).}
\label{alphaexacto}
\end{center}
\end{figure}
\begin{figure}
\begin{center}
\hspace{0.25cm}
   \includegraphics[width=7.5cm]{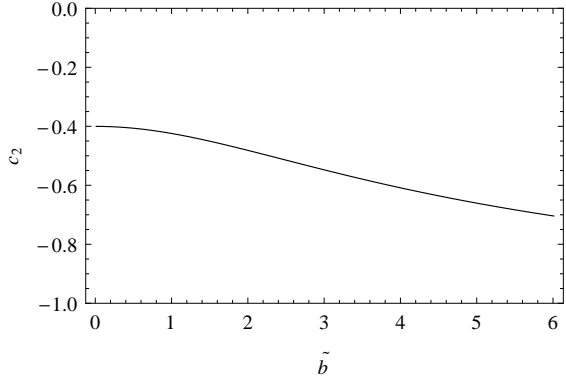}
\caption{Strong deflection limit coefficient $c_2$ as a function of $\tilde{b}$.}
\label{c2plot}
\end{center}
\end{figure}

\noindent The deflection angle is a monotonic decreasing function of $x_0$, as can be seen in Fig. \ref{alphaexacto}. It diverges when $x_0$ approaches to the radius of the photon sphere $x_{ps}=3$, and goes to zero for large $x_0$. When $x_0$ is close enough to $x_{ps}$, the deflection angle is greater than $2\pi$, which means that the photons perform one or more turns around the black hole before emerging towards the observer. This results in two infinite sets of relativistic images, one on each side or the black hole (i.e. at the same side and at the opposite side of the source). To study the situation where the photons pass close to the photon sphere, we adopt the so-called strong deflection limit (Bozza 2002). The integral (\ref{I0}) can be split into two parts
\begin{equation}
I(x_0)=I_D(x_0)+I_R(x_0),
\end{equation}
where $I_D$ yields the leading order term in the divergence of the deflection angle, which is logarithmic; $I_R$ is regular for all values of $x_0$ and can be easily evaluated. It can be shown that in this limit, the deflection angle can be approximated by the simple general form (Bozza 2002)
\begin{equation}
\alpha(u)=-c_1\ln\left(\frac{u}{u_{ps}}-1\right)+c_2+O(u-u_{ps}),
\label{alphasdl}
\end{equation}
with $u=[C(x_0)A^{-1}(x_0)]^{1/2}$ the impact parameter, and $u_{ps}=u(x_{ps})$. The quantities $c_1$ and $c_2$ are the strong deflection limit coefficients, which depend only on the metric functions. Performing the calculations for the regular phantom black hole, we have that $c_1=1$ is a constant, and $c_2$ results (Eiroa et al. 2013)
\begin{equation}
c_2=-\pi+c_R+\ln \frac{\tilde{b}^3\left[-6\tilde{b}+9\pi+\tilde{b}^2\pi-2(9+\tilde{b}^2)\tan^{-1}\left(\frac{3}{\tilde{b}}\right)\right]^2}{(9+\tilde{b}^2)^2\left[2\tilde{b}-3\pi+6\tan^{-1}\left(\frac{3}{\tilde{b}}\right)\right]^3},
\end{equation}
where $c_R=I_R(x_{ps})$ is obtained numerically for each value of the parameter $\tilde{b}$. The coefficient $c_2$ is a decreasing function of $\tilde{b}$, as shown in Fig. \ref{c2plot}.

\begin{figure}
\begin{center}
\hspace{0.25cm}
   \includegraphics[width=7.8cm]{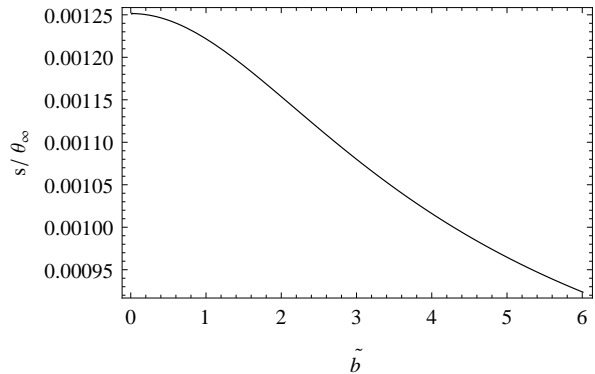}
\caption{Observable $s/\theta_{\infty}$ as a function of $\tilde{b}$.}
\label{stita}
\end{center}
\end{figure}

Adopting a configuration where the black hole ($l$) is situated between the source ($s$) and the observer ($o$), both located in the (asymptotically) flat region, at distances $x\gg x_{h}$, the lens equation is given by (Bozza 2008)
\begin{equation}
\tan\beta=\frac{d_{ol}\sin\theta-d_{ls}\sin(\alpha-\theta)}{d_{os}\cos(\alpha-\theta)},
\end{equation}
where $d_{os}$, $d_{ol}$ and $d_{ls}$, are the observer-source, observer-lens, and lens-source angular diameter distances, respectively; $\beta$ is the angular position of the source, and $\theta$ is the angular position of the image. For highly aligned objects, i.e. $\beta$ and $\theta$ small, the lensing effects are more significant, and the deflection angle, for each set of relativistic images, is close to an even multiple of $\pi$:
\begin{equation}
\alpha=\pm2n\pi\pm\Delta\alpha_n,
\end{equation}
where $n\in\mathbb{N}$, and $0<\Delta\alpha_n\ll 1$. Then, the lens equation takes the simplified form
\begin{equation}
\beta=\theta\mp\frac{d_{ls}}{d_{os}}\Delta\alpha_n.
\label{lens}
\end{equation}
From the lens geometry, we have that $u=d_{ol}\sin\theta\approx d_{ol}\theta$. Replacing this relation in Eq. (\ref{alphasdl}), and using the lens equation (\ref{lens}), it is not difficult to see that the angular position of the $n$-th relativistic image results (Bozza 2002)
\begin{equation}
\theta_n=\pm\theta^{0}_{n}+\frac{\xi_nd_{os}}{d_{ls}}(\beta\mp\theta^{0}_{n}),
\label{thetan}
\end{equation}
where
\begin{equation}
\theta^{0}_{n}=\frac{u_{ps}}{d_{ol}}\left[1+e^{(c_2-2n\pi))/c_1}\right],
\label{theta0}
\end{equation}
and
\begin{equation}
\xi_n=\frac{u_{ps}}{c_1 d_{ol}}e^{(c_2-2n\pi)/c_1}.
\label{xi}
\end{equation}
The magnification of the $n$-th image $\mu_n$ is given by the quotient between the angle subtented by the image and the source:
\begin{equation}
\mu_n=\left|\frac{\beta}{\theta_n}\frac{d\beta}{d\theta_n}\right|^{-1}.
\label{mag}
\end{equation}
Replacing Eq. (\ref{thetan}) in expression (\ref{mag}), within the approximations adopted above, the magnification of each relativistic image takes the form
\begin{equation}
\mu_n=\frac{1}{\beta}\frac{\theta^{0}_{n}\xi_n d_{os}}{d_{ls}}.
\label{mag2}
\end{equation}
The magnifications decrease exponentially with $n$, so the first relativistic image is the brighest one, as can be seen by replacing expressions (\ref{theta0}) and (\ref{xi})  in (\ref{mag2}). These magnitudes can be related with observations by defining the observables:
\begin{equation}
s=\theta_1-\theta_\infty
\end{equation}
and
\begin{equation}
r=\frac{\mu_1}{\sum^{\infty}_{n=2}\mu_n}.
\end{equation}
The angular position $\theta_{\infty}$ is the limiting value where the images approach as $n\rightarrow\infty$, which is an increasing function of $\tilde{b}$ for a given value of $d_{ol}$. The first relativistic image is expected to be resolved from the others since it is the outermost and brightest one. Then, the observable $s$ is defined as the angular separation between the first relativistic image and the others, which are packed together at the limiting value $\theta_{\infty}$. The observable $r$ is the quotient between the flux of the first relativistic image and the flux coming from all the others. In the strong deflection limit, and for high alignment, we obtain that
\begin{equation}
s=\theta_\infty e^{(c_2-2\pi)/c_1}=\theta_\infty e^{c_2-2\pi}
\end{equation}
and
\begin{equation}
r=e^{2\pi/c_1}=e^{2\pi},
\end{equation}
which are functions of $c_1$ and $c_2$. For the phantom black hole, $c_1=1$, so $r$ is a constant. The quotient $s/\theta_{\infty}$ is plotted as a function of $\tilde{b}$ in Fig. \ref{stita}.

\section{Summary and discussion}
In this work, uncharged regular phantom black holes, obtained from a scalar field possessing a negative kinetic term and a potential, were studied as gravitational lenses. The strong deflection limit coefficients $c_1$ and $c_2$ were calculated, which allowed us to obtain analytical expressions for the positions and the magnifications of the relativistic images, and the observables $\theta_{\infty}$, $r$ and $s$. The first relativistic image is the outer one, which is also the brightest, and the others are packed together at the limiting value $\theta_{\infty}$. If the strong deflection limit coefficients can be obtained from observational data, the phantom black holes studied here can be clearly distinguished from the Schwarzschild and vacuum Brans-Dicke solutions, because the values of $c_2$ are different (Bozza 2002; Sarkar et al. 2006).\\

\acknowledgments This work has been supported by CONICET and UBA.

\end{document}